\documentclass[12pt]{iopart}
\usepackage{fancyhdr} 
\usepackage{graphicx}
\usepackage[numbers,sort&compress]{natbib}
\usepackage{indentfirst}
\usepackage{float}
\usepackage{hyperref}
\usepackage{subfigure}
\usepackage{color,xcolor}
\usepackage{soul}
\soulregister\cite7
\soulregister\citep7
\soulregister\citet7
\soulregister\ref7
\soulregister\pageref7

\makeatletter
\@namedef{ver@amsmath.sty}{}
\makeatother
\usepackage{amstext}
\usepackage{fancyhdr}
\begin{document}

\title[Optimal control of complex networks with conformity behavior]{Optimal control of complex networks with conformity behavior}

\author[1]{Zu-Yu Qian$^1$, Cheng Yuan$^1$, Jie Zhou$^1$, Shi-Ming Chen$^1$ and Sen Nie$^1{}^*$}

\address{$^1$ School of Electrical and Automation Engineering, East China Jiaotong University, Nanchang, Jiangxi 330013, China}
\ead{niesen@ecjtu.edu.cn}
\vspace{10pt}
\begin{indented}
\item[]August 2021
\end{indented}

\begin{abstract}
Despite the significant advances in identifying the driver nodes and energy requiring in network control, a framework that incorporates more complicated dynamics remains challenging. Here, we consider the conformity behavior into network control, showing that the control of undirected networked systems with conformity will become easier as long as the number of external inputs beyond a critical point. We find that this critical point is fundamentally determined by the network connectivity. In particular, we investigate the nodal structural characteristic in network control and propose optimal control strategy to reduce the energy requiring in controlling networked systems with conformity behavior. We examine those findings in various synthetic and real networks, confirming that they are prevailing in describing the control energy of networked systems. Our results advance the understanding of network control in practical applications. 
\end{abstract}

\begin{indented}
\item[]{\textbf{Keywords}}: control energy, conformity behavior, optimal control, networked system
\end{indented}

%
%
%
%
%

\section{Introduction}
\noindent Network has been a powerful structure representation of social~\cite{newman2002assortative,albert2002statistical,barabasi2002evolution,newman2003structure,petersen2014reputation}, economic~\cite{mantegna1999introduction,gabaix2003theory}, biological~\cite{jeong2001lethality,rubinov2010complex,yan2017network} and transportation systems~\cite{guimera2005worldwide,jiang2014traffic,wang2017capturing}. A network is composed by a node set and the interactions between nodes are quantified as the link set. In the networks, the node state can be altered by its neighbors, paving the way of controlling them~\cite{liu2011controllability,yuan2013exact,xiang2019advances}. A dynamical system is controllable if it can be driven from any initial state to any desired state with external inputs (driver nodes) within finite time~\cite{rugh1996linear}, and the controllability of networks can be evaluated by the minimal number of inputs to achieve full control. Identifying the minimal driver nodes set of a network can be mapped into determining the maximum matching problem of corresponding bipartite graph~\cite{liu2011controllability}. Recent advances in control mode~\cite{nepusz2012controlling,jia2013control,gao2014target,wang2016controllability,gao2016universal}, optimal driver nodes set~\cite{wang2012optimizing,posfai2013effect,ruths2014control} and control of nonlinear dynamic systems~\cite{cowan2012nodal,cornelius2013realistic} have offered important insights in understanding the controllability of complex networked systems. Moreover, energy requiring is another key characteristic when control a network in practice, which is simultaneously determined by controllability Gramian matrix, control time, initial state and final state~\cite{lewis2012optimal,yan2012controlling,sun2013controllability,nie2018control,yan2015spectrum}. Despite recent studies in control inputs~\cite{wang2017physical,nie2016effect}, control trajectory~\cite{sun2013controllability} and optimal control~\cite{tzoumas2015minimal,chen2016energy,lindmark2018minimum,wang2021optimizing} have extended our understanding in network control, the canonical linear dynamics is too simple to describe the complex networked systems in practice~\cite{pasqualetti2014controllability,nie2020structural}. 

Conformity is an ubiquitous behavior can be observed in natural and social systems, showing important functions in game theory~\cite{boyd1988culture,perc2017statistical,wang2015impact}, human behavior dynamics~\cite{szolnoki2015conformity,wang2010aspiring} and multi-agent systems~\cite{olfati2007consensus,wang2016synchronization}. Wang et al.~\cite{wang2015controlling} introduced the conformity into network control and found that the dynamical networked system with conformity behavior tends to require less number of driver nodes. Chen et al.~\cite{chen2021energy} compared the energy cost for social networks with and without conformity behavior, and estimated the lower and upper bound of energy according to control time. In this paper, we explore the energy requiring in controlling various synthetic and empirical networks with conformity behavior. We find that it is easier to control the conformity-incorporated systems with identical final state as the number of control inputs beyonds the critical point. In particular, we find that nodal characteristic plays vital role in network control and directly driving the hub nodes can yield lower energy than that randomly selected nodes in controlling networks with conformity dynamics. We confirm those findings in several real networks, demonstrating that they are prevailing in describing the control energy of networked systems.

\section{Model}
\noindent The dynamics of a controlled network with $N$ nodes and $M$ external control inputs is
\begin{equation}
\label{eq1}
    \dot{\mathbf{x}}(t)=A\mathbf{x}(t)+B\mathbf{u}(t)
\end{equation}
where $\mathbf{x}(t)=[x_{1}(t),x_{2}(t),\cdots,x_{N}(t)]^{T}$ represents the state of the system at time $t$, $\mathbf{u}(t)=[u_1(t),u_2(t),\cdots,u_M(t)]^T$ is the external input. $A$ is an $N \times N$ adjacency matrix which captures the interactions between nodes. $B$ is an $N \times M$ input matrix to describe how the inputs are imposed on nodes, and $B_{ij}=1$ if input $u_j(t)$ is imposed on node $i$. The system shown in Eq.\ref{eq1} is  non-conformity dynamical system. However, conformity behavior implies that the individual's behavior tends to be accordance with its neighbors’. Therefore its state at the next time will be the average state of its neighbors at present time~\cite{wang2015controlling}
\begin{equation}
\label{eq2}
    x_i(t+1) = \sum_{j=1}^{n_i}x_j(t)/k_i
\end{equation}
where $n_i$ is the number of node $i$'s neighbors, and $k_i=\sum_{j=1}^{N}A$ is the degree of node $i$. Then, the conformity dynamics of a controlled network with $N$ nodes can be formalized as
\begin{equation}
\label{eq3}
    \mathbf{x}(t+1) = K^{-1}A\mathbf{x}(t)+B\mathbf{u}(t)
\end{equation}
where $K^{-1}$ is a diagonal matrix of the reciprocal of the degrees (if $k_i=0$ in the matrix $K$, we set $k_i^{-1}=0$). 

A networked system described by Eq.\ref{eq1} can be driven from any initial state $\mathbf{x}_0$ to any final state $\mathbf{x}_f$ within the time $t\in[t_0,t_f]$ using a finite number of control inputs $u(t)$~\cite{rugh1996linear}. The energy required to control the continuous-time system will be $E(t_f)=\int_{t_0}^{t_f}\mathbf{u}^T(t)\mathbf{u}(t)\mathrm{d}t$, and for discrete-time linear system, it will be $E(\tau_f)=\sum\limits_{\tau=0}^{\tau_f-1}\mathbf{u}^T(\tau)\mathbf{u}(\tau)$. According to classical control theory~\cite{rugh1996linear}, the minimal energy to steer the continuous-time system described by Eq.\ref{eq1} from state $\mathbf{x}_0$ at $t=0$ to state $\mathbf{x}_f$ at time $t=t_f$ is
\begin{equation}
\label{eq4}
    E(t_f)=(\mathbf{x}_f-e^{At_f}\mathbf{x}_0)^TW_c^{-1}(t_f)(\mathbf{x}_f-e^{At_f}\mathbf{x}_0)
\end{equation}
where $W_c(t_f)=\int_0^{t_f}e^{At}BB^Te^{A^Tt}\mathrm{d}t$ is controllability Gramian matrix, and the energy-optimal control input is  $u^*(t)=B^Te^{A^T(t_f-t)}W_c^{-1}(t_f)(\mathbf{x}_f-e^{At_f}\mathbf{x}_0)$.
Meanwhile, for discrete-time linear system, the minimum energy required to steer the system described by Eq.\ref{eq3} from state $\mathbf{x}_0$ at $\tau=0$ to state $\mathbf{x}_f$ at time $\tau=\tau_f$ is
\begin{equation}
\label{eq5}
        E(\tau_f)=(\mathbf{x}_f-A^{\tau_f}\mathbf{x}_0)^TW_c^{-1}(\tau_f)(\mathbf{x}_f-A^{\tau_f}\mathbf{x}_0)
\end{equation}
where $W_c(\tau_f)=\sum\limits_{\tau=0}^{\tau_f-1}A^{\tau_f-\tau-1}BB^T(A^T)^{\tau_f-\tau-1}$ is the controllability Gramian matrix for discrete-time system, and the optimal control input is $u^*(\tau)=B^T{(A^T)}^{\tau_f-\tau-1}W_c^{-1}(\tau_f)(\mathbf{x}_f-A^{\tau_f}\mathbf{x}_0)$. The system is controllable as Gramian matrix is non-singular. As for the controllability framework of discrete-time system is similar to that in continuous-time system~\cite{lewis2012optimal}, we set $
\tau_f\rightarrow\infty$ and the control energy $E (\tau_f\rightarrow\infty)$ to satisfy the controllability Gramian matrix is invertible~\cite{yan2015spectrum}. In addition, to guarantee the stability of system, we add a self-loop $A_{ii}=-(\delta + \sum_{j=1}^{N}A_{ij})$ to each node so that the eigenvalues of adjacency matrix $A$ are all negative values for continuous-time system~\cite{yan2015spectrum}, and $\delta=0.25$ is a small perturbation. For discrete-time system, the value of $\delta$ will be varied to guarantee the  stability of system. Note that, the Gramian matrix of dynamical systems with conformity behavior will be simultaneously determined by the matrix $K^{-1}$, $A$ and $B$, allowing the control energy $E$ be differ from that when conformity is absent.

\section{Results}
\begin{figure}
\centering
\includegraphics[width=12cm]{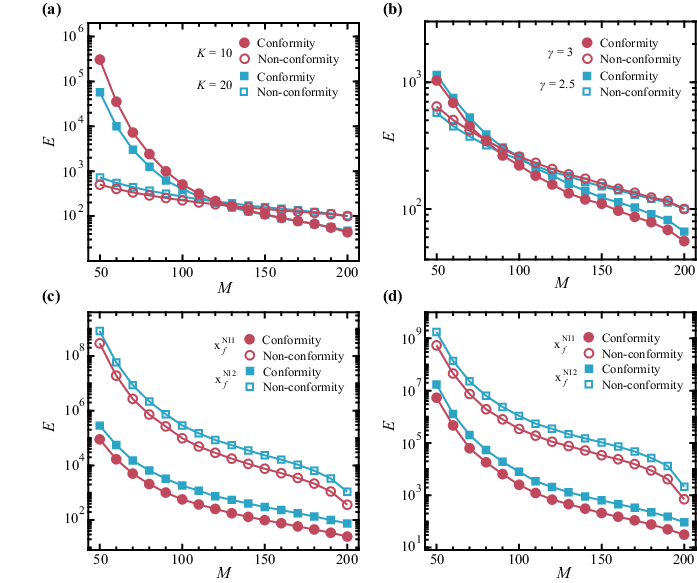}
\caption{\label{fig1} \textbf{Control energy for driving undirected networks as a function of control inputs.} \textbf{(a)} random networks with identical final state, \textbf{(b)} scale-free networks with identical final state, \textbf{(c)} random networks with non-identical final states, and \textbf{(d)} scale-free networks with non-identical final states. Network size $N=200$ and the average degree $K=20$. We set origins as the initial state $\mathbf{x}_0$ of the system and the identical final state $\mathbf{x}_f=[1,1,\cdots,1]^T$.  $\mathbf{x}_{f}^{\rm NI1}$ and $\mathbf{x}_{f}^{\rm NI2}$ are two non-identical final states. For random network with conformity behaviors, we set $\delta=-0.9$ to guarantee the stability of discrete-time system. For scale-free network with conformity behaviors, we add a self-loop $A_{ii}=-0.25$ to each node to make sure the system is stable. For networks without conformity, we set $\delta=0.25$. The driver nodes are randomly selected and each data point is the mean of 100 independent realizations.}
\end{figure}

\noindent To reveal whether conformity behavior can fundamentally change the network control effort, we calculate the control energies of canonical dynamical system (Eq.\ref{eq1}) and conformity dynamical system (Eq.\ref{eq2}), respectively. We set $M$ as the external inputs, finding that the control energy of both systems can be efficiently decreased through imposing more inputs (Fig.\ref{fig1}). Interestingly, energy requiring difference between conformity and canonical dynamical systems displays a transition trajectory when systems toward identical final state, implying that those conformity behaviors between nodes are not always facilitating the network control. This finding is robust to average degree and degree distribution (see Fig.\ref{fig1}a for undirected Erd$\ddot{o}$s-R$\acute{e}$nyi random graph~\cite{erdHos1960evolution}, and Fig.\ref{fig1}b for scale-free networks~\cite{goh2001universal}). Yet, the critical point $M_{\text c}$ of energy difference between canonical and conformity systems do depend on the average degree.

Though synchronization is prevalent in nature, in which individuals toward to the identical final state, there are also some cases for individuals reaching different final states. Such as in cooperative control of multi-agent systems, the individuals finally with non-identical states to fulfill the task. In order to compare the energy requiring for controlling networks with different final states, we take two different non-identical final states into consideration. (i) $\mathbf{x}_{f}^{\rm NI1}$, where the final state of each node is drawn from the uniform distribution $x_i(t_{f}) \sim \mathbf{U}(0,1)$. (ii) $\mathbf{x}_{f}^{\rm NI2}$, where the final state of each node is drawn from the uniform distribution $x_i(t_{f}) \sim \mathbf{U}(0,\sqrt{3})$. To guarantee the fairness of comparison between identical and non-identical final state, the Euclidean distances from $\mathbf{x}_0$ to the two final states $\mathbf{x}_{f}^{\rm NI2}$ and $\mathbf{x}_f$(identical final state) are the same. For random graph in Fig.\ref{fig1}c and scale-free network in Fig.\ref{fig1}d, the  control energy for driving networks to distinct final states are compared. The result shows it is still easier to control networks with conformity behaviors. However, the energy transition between conformity-incorporated and canonical dynamical systems disappears. Fig.\ref{fig1}c-d shows the control energy for network with final state $\mathbf{x}_{f}^{\rm NI1}$ is lower than that with $\mathbf{x}_{f}^{\rm NI2}$. Meanwhile, with the same Euclidean distance from initial state to final state, the energy requiring for controlling network with identical final state $\mathbf{x}_f$ is lower than that with non-identical final state $\mathbf{x}_{f}^{\rm NI2}$.

\begin{figure}
\centering
\includegraphics[width=12cm]{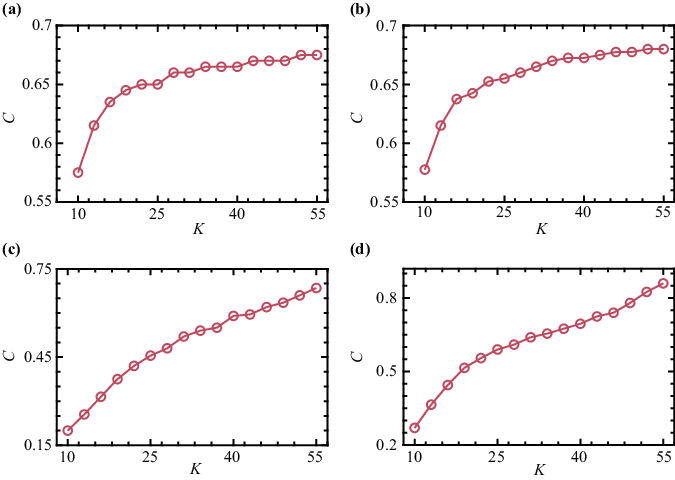}
\caption{\label{fig2} \textbf{Critical point fraction of inputs as a function of average degree.} \textbf{(a-b)} random networks, \textbf{(c-d)} scale-free networks. The network size for \textbf{(a)} and \textbf{(b)} are $N=200$ and $N=400$, the degree exponent for \textbf{(c)} and \textbf{(d)} are $\gamma=2.5$ and $\gamma=3$. Each data point is the mean of 100 independent realizations.}
\end{figure}

We further investigate the critical input point $C=M_{\text c}/N$ in terms of average degree, finding that critical point increases monotonically when the network connectivity is increased (see Fig.\ref{fig2}). For random graphs (Fig.\ref{fig2}a,b), the critical point will finally reach around 0.67 while for scale-free networks (see Fig\ref{fig2}c,d), the value keeps increasing with average degree. This suggests that the efficacy of conformity behavior in reducing the control energy is fundamentally determined by the network connectivity. For sparse networks, critical point $C$ in scale-free networks are much smaller than that in random networks. It is known that conformity propels the state of nodes to be consistent with their neighbors, thus it has a big effect on sparse and heterogeneous networks, in which connections between nodes are less.

\begin{figure}
\centering
\includegraphics[width=12cm,height=6cm]{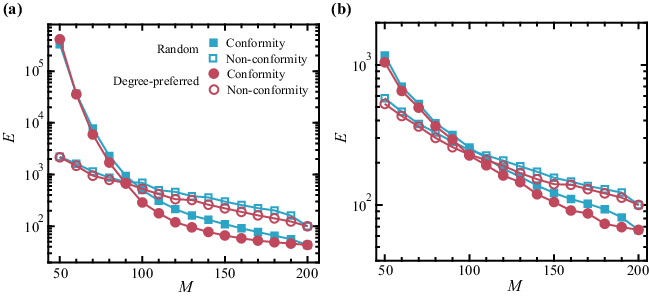}
\caption{\label{fig3} \textbf{Control energy for driving undirected networks with different strategies.} \textbf{(a)} random networks, \textbf{(b)} scale-free networks. Average degree $K=10$, degree exponent $\gamma=3$ and network size $N=200$. The statistics and parameters not shown are the same as those used in Fig.\ref{fig1}. Each data point is the mean of 100 independent realizations.}
\end{figure}

Previous studies demonstrated that the characteristic of driver nodes set might intrinsically impact the energy requiring. Therefore, we develop two control strategies of driver nodes selection, examining the energy effort in control of networked systems (see Fig.\ref{fig3}). (i) Random: a fraction of driver nodes was randomly selected. (ii) Degree-preferred: selecting driver nodes according to the descending order of their degree. We find that control energy by applying those two strategies shows the similar transition trajectories as observed in the Fig.\ref{fig1}, e.g. for small $M$, imposing inputs on those hub nodes cannot efficiently reduce the energy requiring. When more external inputs are introduced, it is easier to control the networked systems by driving hub nodes. Compared with the energy requiring to control networks with conformity behavior, there is no obvious difference of energy between two strategies for controlling both Erd$\ddot{o}$s-R$\acute{e}$nyi random graphs and scale-free networks without conformity. The results indicate the significance of structural characteristic makes the hub nodes prominent in controlling networks, which directly driving them could reduce the energy.

\begin{figure}
\centering
\includegraphics[width=12cm,height=6cm]{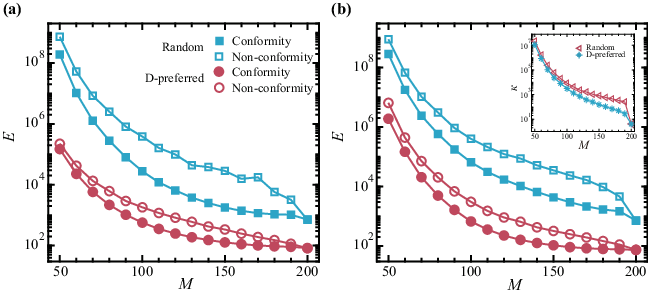}
\caption{\label{fig4} \textbf{Control energy for driving directed networks with different strategies.} \textbf{(a)} random networks, \textbf{(b)} scale-free networks. Inset: condition number $\kappa$ of Gramian matrix of directed scale-free networks with conformity dynamics as a function of control inputs $M$. Network size $N=200$, $K=10$ and degree exponent $\gamma_\text{in}=\gamma_\text{out}=3$. The statistics and parameters not shown are the same as those used in Fig.\ref{fig1}. Each data point is the mean of 100 independent realizations.}
\end{figure}

We now turn to the controlling of conformity dynamics on directed networks. The driver nodes are also selected according to the following two strategies: (i) Random: a fraction of driver nodes was randomly selected. (ii) D-preferred: here a fraction of driver nodes was chosen based on the descending order of the difference between their out-degrees and in-degrees $k_\text{out}-k_\text{in}$. We find that, with the same driver node selection strategy, control energy for driving directed networked systems with conformity are always lower than that for systems without conformity. There is no energy transition with increased inputs, which is different from the results for undirected networks. On the other hand, imposing external inputs to the hub nodes can efficiently reduce the energy requiring, both for random graph (Fig.\ref{fig4}a) and scale-free network (Fig.\ref{fig4}b). Moreover, the energy difference between two strategies are larger than that for undirected networks, and this phenomenon exists both in canonical dynamical system and conformity-incorporated system.

\begin{figure}
\centering
\includegraphics[width=12cm]{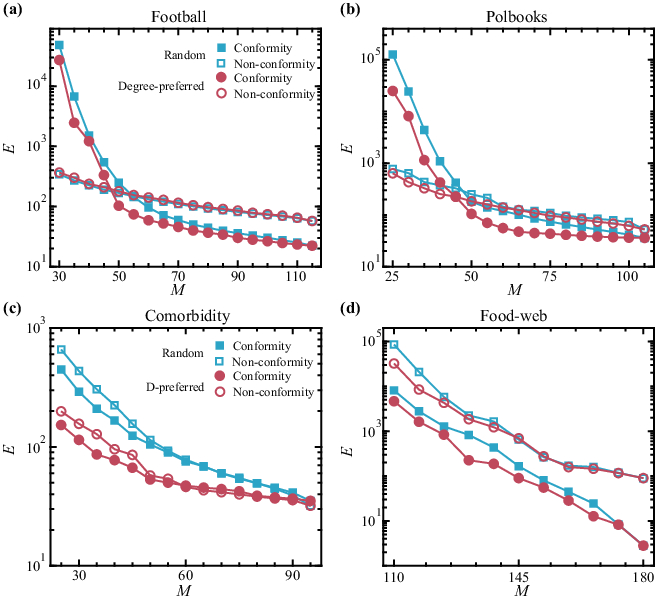}
\caption{\label{fig5} \textbf{Control energy for driving four real networks as a function of control inputs.} \textbf{(a)} Football, \textbf{(b)} Politics books, \textbf{(c)} Human comorbidity, \textbf{(d)} Food-web. Each data point is the mean of 100 independent realizations.}
\end{figure}

Finally, we examine the impact of conformity behavior as well as the driver node set on the control of various real networks. To achieve this, we select two undirected networks: Football  (with 115 nodes and 616 links)~\cite{girvan2002community}, Politics books (with 105 nodes and 441 links)~\cite{Krebs2014}; and two directed networks: Human comorbidity (with 95 nodes and 8930 links)~\cite{alexander2020architecture}, Food-web (with 180 nodes and 1577 links)~\cite{zander2011food}. We find that transition behavior of control energy between conformity and non-conformity dynamical systems can be observed in some real networks (see Fig.\ref{fig5}a,b). Dynamical systems without conformity displays advantage only as the number of control inputs is very limited. Besides, for other real networks (see Fig.\ref{fig5}c,d), the energy requiring are still lower as small number of inputs for networks with conformity. Meanwhile, for networks with conformity dynamics, driving the hub nodes or the nodes with large value of $k_\text{out}-k_\text{in}$ can yield lower control energy than that of randomly selected nodes.

\section{Discussion}
\noindent We explore the influence of conformity behavior on control energy for both synthetic and real networks. By comparing the energy for driving networks with and without conformity behavior, we note that for undirected networks with identical final state, comparison of energy varied as the number of control inputs beyonds the critical point, and it is easier to control networks incorporated with conformity dynamics as a large number of inputs. We also find the critical point is tied with the network connectivity. For directed networks with conformity, the energy requiring are always lower than that without conformity. Further, this paper investigates how the nodal characteristics affects the control energy. Different control strategies based on structural character are chosen for undirected and directed networks, respectively. The results show it is easier to control undirected networks with conformity behavior by driving the hub nodes, and controlling the nodes with large difference between out-degrees and in-degrees for directed network is beneficial to control energy. Finally, the results are verified in real networks. Our work extends understanding of control energy for networked system with dynamics, which advances the application of network science in practical control.

\section*{Acknowledgments}

\noindent This work was supported by the National Natural Science Foundation of China (No.61763013), The Natural Science Foundation of Jiangxi Province (No.20202BABL212008), and the Jiangxi Provincial Postdoctoral Preferred Project of China (No.2017KY37). The authors wish to thank Dr. Xu-Wen Wang for valuable comments and discussion throughout the development of this work.

\bibliographystyle{iopart-num.bst}
\bibliography{iopart-num.bib}

\providecommand{\noopsort}[1]{}\providecommand{\singleletter}[1]{#1}%
\providecommand{\newblock}{}
\begin{thebibliography}{10}
\expandafter\ifx\csname url\endcsname\relax
  \def\url#1{{\tt #1}}\fi
\expandafter\ifx\csname urlprefix\endcsname\relax\def\urlprefix{URL }\fi
\providecommand{\eprint}[2][]{\url{#2}}

\bibitem{newman2002assortative}
Newman M~E 2002 {\em Phys. Rev. L\/} {\bf 89} 208701

\bibitem{albert2002statistical}
Albert R and Barab{\'a}si A~L 2002 {\em Rev. Mod. Phys.\/} {\bf 74} 47

\bibitem{barabasi2002evolution}
Barab{\^a}si A~L, Jeong H, N{\'e}da Z, Ravasz E, Schubert A and Vicsek T 2002
  {\em Physica A\/} {\bf 311} 590--614

\bibitem{newman2003structure}
Newman M~E 2003 {\em SIAM Rev\/} {\bf 45} 167--256

\bibitem{petersen2014reputation}
Petersen A~M, Fortunato S, Pan R~K, Kaski K, Penner O, Rungi A, Riccaboni M,
  Stanley H~E and Pammolli F 2014 {\em Proc. Natl Acad. Sci. USA\/} {\bf 111}
  15316--15321

\bibitem{mantegna1999introduction}
Mantegna R~N and Stanley H~E 1999 {\em Introduction to econophysics:
  correlations and complexity in finance\/} (Cambridge University Press)

\bibitem{gabaix2003theory}
Gabaix X, Gopikrishnan P, Plerou V and Stanley H~E 2003 {\em Nature\/} {\bf
  423} 267--270

\bibitem{jeong2001lethality}
Jeong H, Mason S~P, Barab{\'a}si A~L and Oltvai Z~N 2001 {\em Nature\/} {\bf
  411} 41--42

\bibitem{rubinov2010complex}
Rubinov M and Sporns O 2010 {\em Neuroimage\/} {\bf 52} 1059--1069

\bibitem{yan2017network}
Yan G, V{\'e}rtes P~E, Towlson E~K, Chew Y~L, Walker D~S, Schafer W~R and
  Barab{\'a}si A~L 2017 {\em Nature\/} {\bf 550} 519--523

\bibitem{guimera2005worldwide}
Guimera R, Mossa S, Turtschi A and Amaral L~N 2005 {\em Proc. Natl Acad. Sci.
  USA\/} {\bf 102} 7794--7799

\bibitem{jiang2014traffic}
Jiang R, Hu M~B, Zhang H, Gao Z~Y, Jia B, Wu Q~S, Wang B and Yang M 2014 {\em
  PloS One\/} {\bf 9} e94351

\bibitem{wang2017capturing}
Wang X, Jiang R, Li L, Lin Y, Zheng X and Wang F~Y 2017 {\em IEEE Trans. Int.
  Tra. Sys.\/} {\bf 19} 910--920

\bibitem{liu2011controllability}
Liu Y~Y, Slotine J~J and Barab{\'a}si A~L 2011 {\em Nature\/} {\bf 473}
  167--173

\bibitem{yuan2013exact}
Yuan Z, Zhao C, Di Z, Wang W~X and Lai Y~C 2013 {\em Nat. Commun.\/} {\bf 4}
  1--9

\bibitem{xiang2019advances}
Xiang L, Chen F, Ren W and Chen G 2019 {\em IEEE Circuits Syst. Mag.\/} {\bf
  19} 8--32

\bibitem{rugh1996linear}
Rugh W~J 1996 {\em Linear system theory\/} (Prentice-Hall, Inc.)

\bibitem{nepusz2012controlling}
Nepusz T and Vicsek T 2012 {\em Nat. Phys.\/} {\bf 8} 568--573

\bibitem{jia2013control}
Jia T and Barab{\'a}si A~L 2013 {\em Sci. Rep.\/} {\bf 3} 1--6

\bibitem{gao2014target}
Gao J, Liu Y~Y, D'souza R~M and Barab{\'a}si A~L 2014 {\em Nat. Commun.\/} {\bf
  5} 1--8

\bibitem{wang2016controllability}
Wang L, Chen G, Wang X and Tang W~K 2016 {\em Automatica\/} {\bf 69} 405--409

\bibitem{gao2016universal}
Gao J, Barzel B and Barab{\'a}si A~L 2016 {\em Nature\/} {\bf 530} 307--312

\bibitem{wang2012optimizing}
Wang W~X, Ni X, Lai Y~C and Grebogi C 2012 {\em Phys. Rev. E\/} {\bf 85} 026115

\bibitem{posfai2013effect}
P{\'o}sfai M, Liu Y~Y, Slotine J~J and Barab{\'a}si A~L 2013 {\em Sci. Rep.\/}
  {\bf 3} 1--7

\bibitem{ruths2014control}
Ruths J and Ruths D 2014 {\em Science\/} {\bf 343} 1373--1376

\bibitem{cowan2012nodal}
Cowan N~J, Chastain E~J, Vilhena D~A, Freudenberg J~S and Bergstrom C~T 2012
  {\em PloS One\/} {\bf 7} e38398

\bibitem{cornelius2013realistic}
Cornelius S~P, Kath W~L and Motter A~E 2013 {\em Nat. Commun.\/} {\bf 4} 1--9

\bibitem{lewis2012optimal}
Lewis F~L, Vrabie D and Syrmos V~L 2012 {\em Optimal control\/} (John Wiley \&
  Sons)

\bibitem{yan2012controlling}
Yan G, Ren J, Lai Y~C, Lai C~H and Li B 2012 {\em Phys. Rev. Lett.\/} {\bf 108}
  218703

\bibitem{sun2013controllability}
Sun J and Motter A~E 2013 {\em Phys. Rev. Lett.\/} {\bf 110} 208701

\bibitem{nie2018control}
Nie S, Stanley H~E, Chen S~M, Wang B~H and Wang X~W 2018 {\em Sci. Rep.\/} {\bf
  8} 1--8

\bibitem{yan2015spectrum}
Yan G, Tsekenis G, Barzel B, Slotine J~J, Liu Y~Y and Barab{\'a}si A~L 2015
  {\em Nat. Phys.\/} {\bf 11} 779--786

\bibitem{wang2017physical}
Wang L~Z, Chen Y~Z, Wang W~X and Lai Y~C 2017 {\em Sci. Rep.\/} {\bf 7} 1--14

\bibitem{nie2016effect}
Nie S, Wang X~W, Wang B~H and Jiang L~L 2016 {\em Sci. Rep.\/} {\bf 6} 1--9

\bibitem{tzoumas2015minimal}
Tzoumas V, Rahimian M~A, Pappas G~J and Jadbabaie A 2015 {\em IEEE Trans.
  Contr. Netw. Syst.\/} {\bf 3} 67--78

\bibitem{chen2016energy}
Chen Y~Z, Wang L~Z, Wang W~X and Lai Y~C 2016 {\em R. Soc. Open Sci.\/} {\bf 3}
  160064

\bibitem{lindmark2018minimum}
Lindmark G and Altafini C 2018 {\em Sci. Rep.\/} {\bf 8} 1--14

\bibitem{wang2021optimizing}
Wang X and Xiang L 2021 {\em Complexity\/} {\bf 2021} 6657307

\bibitem{pasqualetti2014controllability}
Pasqualetti F, Zampieri S and Bullo F 2014 {\em IEEE Trans. Contr. Netw.
  Syst.\/} {\bf 1} 40--52

\bibitem{nie2020structural}
Nie S, Stanley H~E, Chen S~M, Wang B~H and Wang X~W 2020 {\em Europhys.
  Lett.\/} {\bf 130} 58002

\bibitem{boyd1988culture}
Boyd R and Richerson P~J 1988 {\em Culture and the evolutionary process\/}
  (University of Chicago Press)

\bibitem{perc2017statistical}
Perc M, Jordan J~J, Rand D~G, Wang Z, Boccaletti S and Szolnoki A 2017 {\em
  Phys. Rep.\/} {\bf 687} 1--51

\bibitem{wang2015impact}
Wang X~W, Wang Z, Nie S, Jiang L~L and Wang B~H 2015 {\em Appl. Math.
  Comput.\/} {\bf 250} 848--853

\bibitem{szolnoki2015conformity}
Szolnoki A and Perc M 2015 {\em J. R. Soc., Interface\/} {\bf 12} 20141299

\bibitem{wang2010aspiring}
Wang Z and Perc M 2010 {\em Phys. Rev. E\/} {\bf 82} 021115

\bibitem{olfati2007consensus}
Olfati-Saber R, Fax J~A and Murray R~M 2007 {\em Proc. IEEE\/} {\bf 95}
  215--233

\bibitem{wang2016synchronization}
Wang L and Chen G 2016 {\em Chaos\/} {\bf 26} 094809

\bibitem{wang2015controlling}
Wang X~W, Nie S, Wang W~X and Wang B~H 2015 {\em Europhys. Lett.\/} {\bf 111}
  68004

\bibitem{chen2021energy}
Chen H and Yong E~H 2021 {\em arXiv:2101.03828\/}

\bibitem{erdHos1960evolution}
Erd{\H{o}}s P and R{\'e}nyi A 1960 {\em Publ. Math. Inst. Hung. Acad. Sci\/}
  {\bf 5} 17--60

\bibitem{goh2001universal}
Goh K~I, Kahng B and Kim D 2001 {\em Phys. Rev. Lett.\/} {\bf 87} 278701

\bibitem{girvan2002community}
Girvan M and Newman M~E 2002 {\em Proc. Natl Acad. Sci. USA\/} {\bf 99}
  7821--7826

\bibitem{Krebs2014}
Krebs V unpublished \url{http://www.orgnet.com/}

\bibitem{alexander2020architecture}
Alexander-Bloch A~F, Raznahan A, Shinohara R~T, Mathias S~R, Bathulapalli H,
  Bhalla I~P, Goulet J~L, Satterthwaite T~D, Bassett D~S, Glahn D~C {\em
  et~al.\/} 2020 {\em Proc. R. Soc. A\/} {\bf 476} 20190790

\bibitem{zander2011food}
Zander C~D, Josten N, Detloff K~C, Poulin R, McLaughlin J~P and Thieltges D~W
  2011 {\em Ecology\/} {\bf 92} 2007--2007

\end{thebibliography}
\end{document}